\documentclass[letters,fleqn,usenatbib]{mnras}
\usepackage{mathptmx}

\usepackage[T1]{fontenc}
\usepackage{ae,aecompl}

\usepackage[latin1]{inputenc}
\usepackage{multirow}
\usepackage{graphicx} 
\usepackage{multicol}
\usepackage{float}
\usepackage{footnote}
\usepackage{amssymb} 	
\usepackage[figuresright]{rotating}
\usepackage{amsmath} 	
\usepackage{url}
\usepackage{rotating}
\usepackage{caption}

\usepackage{subcaption} 
\usepackage[export]{adjustbox}

\title[The filamentary nebula in NGC\~1275]{Revealing the velocity structure of the filamentary nebula\\
 in NGC 1275 in its entirety}
\author[M. Gendron-Marsolais et al.]{M. Gendron-Marsolais$^{1}$\thanks{E-mail: marie-lou@astro.umontreal.ca}, 
J. Hlavacek-Larrondo$^{1}$,
T. B. Martin$^{2}$, 
L. Drissen$^{2,3}$, 
\newauthor M. McDonald$^{4}$, 
A. C. Fabian$^{5}$,
A. C. Edge$^{6}$, 
S. L. Hamer$^{7}$,
B. McNamara$^{8,9}$ 
\newauthor and G. Morrison$^{10}$
\\
$^{1}$D\'epartement de Physique, Universit\'e de Montr\'eal, Montr\'eal (Qu\'ebec), QC H3C 3J7, Canada\\
$^{2}$D\'epartement de physique, de g\'enie physique et d'optique, Universit\'e Laval, 1045 avenue de la m\'edecine, Qu\'ebec (Qu\'ebec), G1V 0A6, Canada \\
$^{3}$Department of Physics and Astronomy, University of Hawaii at Hilo, 200 W Kawili St., Hilo, HI, USA 9672 \\
$^{4}$Kavli Institute for Astrophysics and Space Research, MIT, Cambridge, MA 02139, USA \\
$^{5}$Institute of Astronomy, University of Cambridge, Madingley Road, Cambridge CB3 0HA, UK\\
$^{6}$Centre for Extragalactic Astronomy, Department of Physics, Durham University, Durham DH1 3LE, UK\\
$^{7}$CRAL, Observatoire de Lyon, CNRS, Universit\'e  Lyon 1, 9 Avenue Charles Andr\'e, F-69561 Saint Genis-Laval, France\\
$^{8}$Department of Physics and Astronomy, University of Waterloo, Waterloo, ON, N2L 3G1, Canada \\
$^{9}$Perimeter Institute for Theoretical Physics, Waterloo, ON, N2L 2Y5, Canada\\
$^{10}$LBT Observatory, University of Arizona, 933 N. Cherry Ave, Room 552, Tucson, AZ 85721 U.S.A.\\}

\vspace{-30pt}
\date{Accepted XXX. Received YYY; in original form ZZZ}

\pubyear{2017}

\begin{document}
\label{firstpage}
\pagerange{\pageref{firstpage}--\pageref{lastpage}}
\maketitle



\begin{abstract}
We have produced for the first time a detailed velocity map of the giant filamentary nebula surrounding NGC 1275, the Perseus cluster's brightest galaxy, and revealed a previously unknown rich velocity structure across the entire nebula. 
We present new observations of the low-velocity component of this nebula with the optical imaging Fourier transform spectrometer SITELLE at CFHT.  
With its wide field of view ($\sim$11'$\times$11'), SITELLE is the only integral field unit spectroscopy instrument able to cover the 80 kpc$\times$55 kpc (3.8'$\times$2.6') large nebula in NGC 1275.
Our analysis of these observations shows a smooth radial gradient of the [N II]$\lambda$6583/$\text{H} \alpha$ line ratio, suggesting a change in the ionization mechanism and source across the nebula, while the dispersion profile shows a general decrease with increasing distance from the AGN at up to $\sim 10$ kpc. 
The velocity map shows no visible general trend or rotation, indicating that filaments are not falling uniformly onto the galaxy, nor being pulled out from it. 
Comparison between the physical properties of the filaments and Hitomi measurements of the X-ray gas dynamics in Perseus are also explored.
\end{abstract}


\section{Introduction}\label{Introduction}
\vspace{-5pt}
The central dominant galaxy of the Perseus galaxy cluster, NGC 1275, is surrounded by a giant filamentary emission-line nebula. 
Such nebulae, with H$_\alpha$ luminosities as high as several 10$^{42}$ erg$/$s, are not rare among clusters having peaked X-ray surface brightness distributions like Perseus, known as cool core clusters (e.g. \citealt{crawford_rosat_1999}). However, the filamentary nebula in NGC 1275 extends over 80 kpc$\times$55 kpc (3.8'$\times$2.6') and is therefore among the largest known in any cluster (e.g. \citealt{mcdonald_optical_2012,hamer_optical_2016}). 
The origin of these nebulae (residual cooling flow, merger gas accretion or dragged gas) and source of ionization (heat conduction from the ICM, shocks or turbulent mixing) are not yet clear. The ionization source does not appear to be related to star formation as the line ratios are different from those in H II regions (e.g. \citealt{kent_ionization_1979}). These nebulae therefore constitute an active area of research for our understanding of how phenomena such as shocks heat and ionize their surrounding medium.

Being the cluster's brightest galaxy (BCG) in Perseus, NGC 1275 resides in a complex environment, both internally perturbed by the nuclear outbursts of its active galactic nuclei (AGN) and externally affected by interactions with its surrounding environment. 
As the brightest cluster in the X-ray sky \citep{forman_observations_1972}, it has been observed across all the electromagnetic spectrum, revealing a variety of structures. X-ray observations of the intracluster medium (ICM) have shown a succession of cavities created by the jets of the central supermassive black hole, pushing away the cluster gas and leaving buoyantly rising bubbles filled with radio emission (e.g. \citealt{fabian_wide_2011}).

First observed by \cite{minkowski_optical_1957} and \cite{lynds_improved_1970},  the nebula surrounding NGC 1275 consists of two distinct components: a high-velocity system ($\sim 8200$ km s$^{-1}$, HV) corresponding to a foreground galaxy, and a low-velocity system ($\sim 5200$ km s$^{-1}$, LV) associated with NGC 1275. 
HST observations of the LV system have revealed a thread-like filamentary composition, some only 70 pc wide and 6 kpc long \citep{fabian_magnetic_2008}. 
The brighter filaments have soft X-ray counterparts \citep{fabian_relationship_2003} and Karl G. Jansky Very Large Array 230-470 MHz observations show a spur of emission in the direction of the northern filament \citep{gendron-marsolais_deep_2017}.
Cold molecular gas are associated with some of the filaments, e.g. CO \citep{salome_cold_2006,ho_multiple_2009,salome_very_2011} and $\text{H}_2$ \citep{lim_molecular_2012}.

The nebula was imaged by \cite{conselice_nature_2001} in its full extent with high-resolution imaging, integral field and long-slit spectroscopy (WIYN \& KPNO).
The authors produced a first velocity map of the central $\sim 45 \arcsec$ (16 kpc), revealing evidence for rotation, and suggested that the filaments were being formed through compression of the hot ICM by the AGN outflows of NGC 1275. 
Further observations from the Gemini Multi-Object Spectrograph along six slits aligned with 2-3 filaments showed evidence of outflowing gas and flow patterns \citep{hatch_origin_2006}.
Overall, this suggests that these filamentary nebulae could be formed by gas being dragged out from the rise of AGN radio bubbles in the ICM and stabilized by magnetic fields \citep{fabian_relationship_2003,hatch_origin_2006,fabian_magnetic_2008}. This is further supported by the presence of a horseshoe-shaped filament, bending behind the North-West outer cavity, similarly to the toroidal flow pattern trailing behind a buoyant gas bubble in a liquid. 
Under this assumption, the loop-like X-ray structure extending at the end of the northern filament  would then be a fallback of gas dragged out to the north by previously formed bubbles \citep{fabian_wide_2011}.
However, the $\text{H} \alpha$ emission found in several cool core clusters' BCGs is delimited within their cooling radius and a strong correlation has been found between $\text{H} \alpha$ luminosity and the X-ray cooling flow rate of the host cluster \citep{mcdonald_origin_2010}. This suggests that the ionized gas may be linked to the ICM and a radially infalling cooling flow model is favoured.

NGC 1275 is one of the richest nebulae to study due to its proximity and the complexity of its structures. 
In this article, we present new observations of NGC 1275 obtained with SITELLE, a new optical imaging Fourier transform spectrometer at Canada-France-Hawaii Telescope (CFHT).
Unlike previous IFU observations of NGC 1275, its wide field of view ($11' \times 11'$) covers the large nebula in its entirety.
To directly compare our results with \cite{hitomi_collaboration_atmospheric_2017}, we adopt a redshift of $z = 0.017284$ for NGC 1275, corresponding to an angular scale of 21.2 kpc arcmin$^{-1}$. This corresponds to a luminosity distance of 75.5 Mpc, assuming $H_{0} = 69.6 \text{ km s}^{-1} \text{Mpc}^{-1}$, $\Omega_{\rm M} = 0.286$ and $\Omega_{\rm vac} = 0.714$.


\vspace{-22pt}
\section{Data reduction and analysis}\label{Data reduction}
\vspace{-5pt}

\begin{figure}
\hspace{-28pt}
\vspace{-10pt}
\centering
\includegraphics[width=0.8\columnwidth]{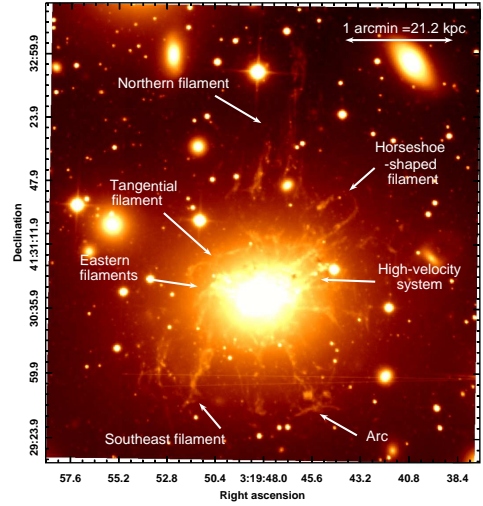}
\caption{Mean integrated flux SN3 filter image centered on NGC 1275. The HV system and the different filaments of the LV system are identified.}
\label{fig:deep_frame}
\vspace{-17pt}
\end{figure}

NGC 1275 was observed  in January 2016 with the optical imaging Fourier transform spectrometer SITELLE at CFHT during Queued Service Observations 16BQ12 in science verification mode (PI G. Morrison) with the SN3 filter ($> 90\%$ transmission from 647-685 nm) for 2.14h (308 exposures of 25 seconds, $R=1800$). 
SITELLE is a Michelson interferometer with a large field of view ($11 \arcmin \times 11 \arcmin$, compare to $1\arcmin\times1\arcmin$ for MUSE and up to $8\arcsec\times8\arcsec$ for SINFONI) equipped with two E2V detectors of $2048 \times 2064$ pixels, resulting in a spatial resolution of $0.321 \arcsec \times 0.321 \arcsec$.
These observations were centered at RA 03h19m53.19s and DEC $+41^{\circ}33\arcmin51.0\arcsec$, offset by about $3 \arcmin$ from NGC 1275.
The data reduction and calibration of these observations were conducted using the SITELLE's software ORCS (version 3.1.2, \citealt{martin_orbs_2015} \footnote{\url{https://github.com/thomasorb/orcs}}).
Five emission lines are resolved in these observations: [N II]$\lambda$6548, $\text{H} \alpha$, [N II]$\lambda$6584, [S II]$\lambda$6716 and [S II]$\lambda$6731. 
Details of the wavelength, astrometric and photometric calibration followed are described in \cite{martin_sitelle_2018}.
The OH sky lines velocities were fitted with an optical model of the interferometer in most regions of the cube with the function \textsc{SpectralCube.map\char`_sky\char`_velocity()} and the resulting wavelength corrections for instrumental flexures were applied to the cube using \textsc{SpectralCube.correct\char`_wavelength()} \citep{martin_sitelle_2018}.
The mean integrated flux SN3 filter image centered on NGC 1275 is shown on figure \ref{fig:deep_frame}.


\begin{figure*}
	\centering
    \begin{subfigure}[t]{0.35\textwidth}
    		\centering
    		\includegraphics[height=7.6cm]{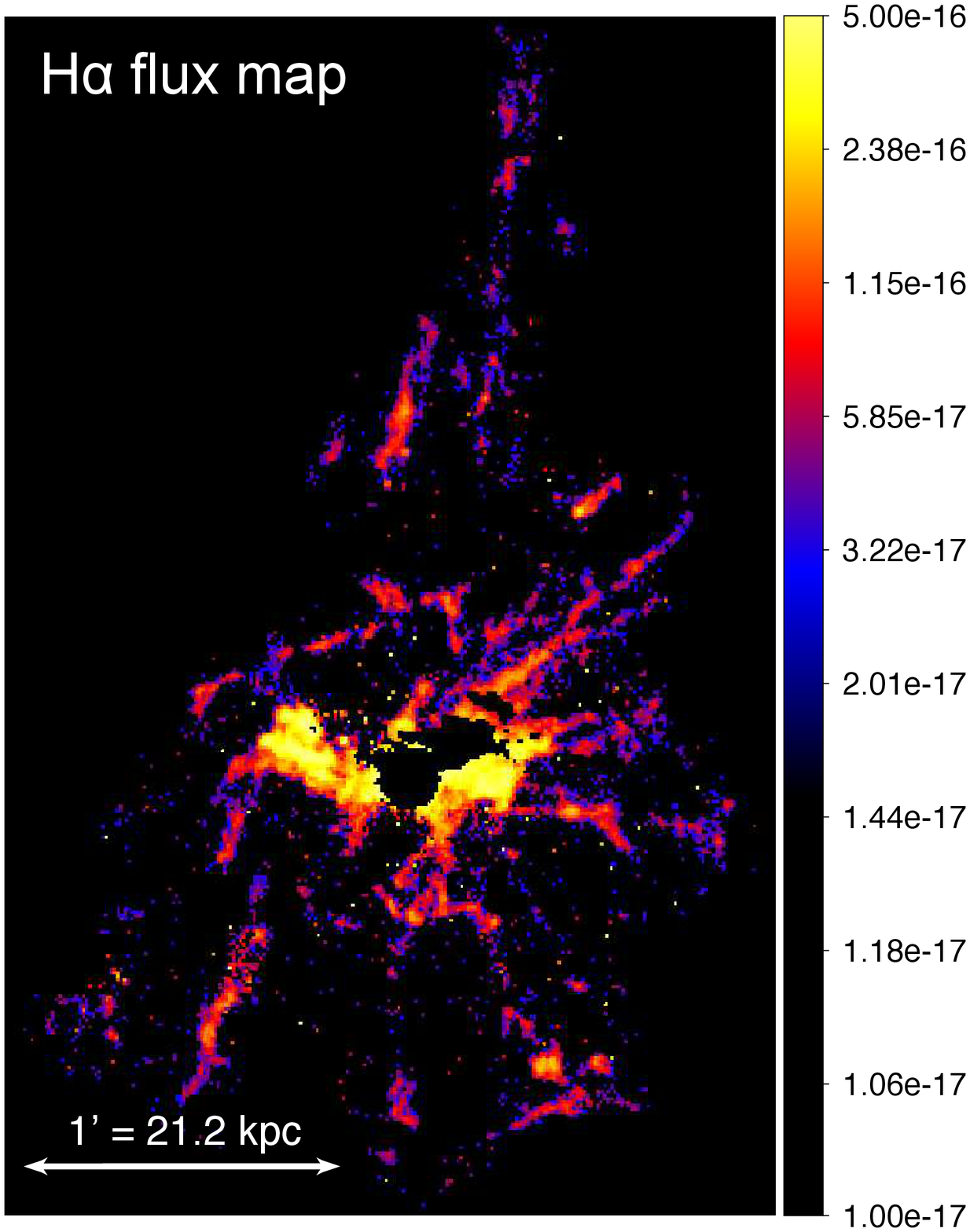}
    		\vspace{-10pt}
    \end{subfigure}
    \begin{subfigure}[t]{0.32\textwidth}
    		\centering
    		\includegraphics[height=7.6cm]{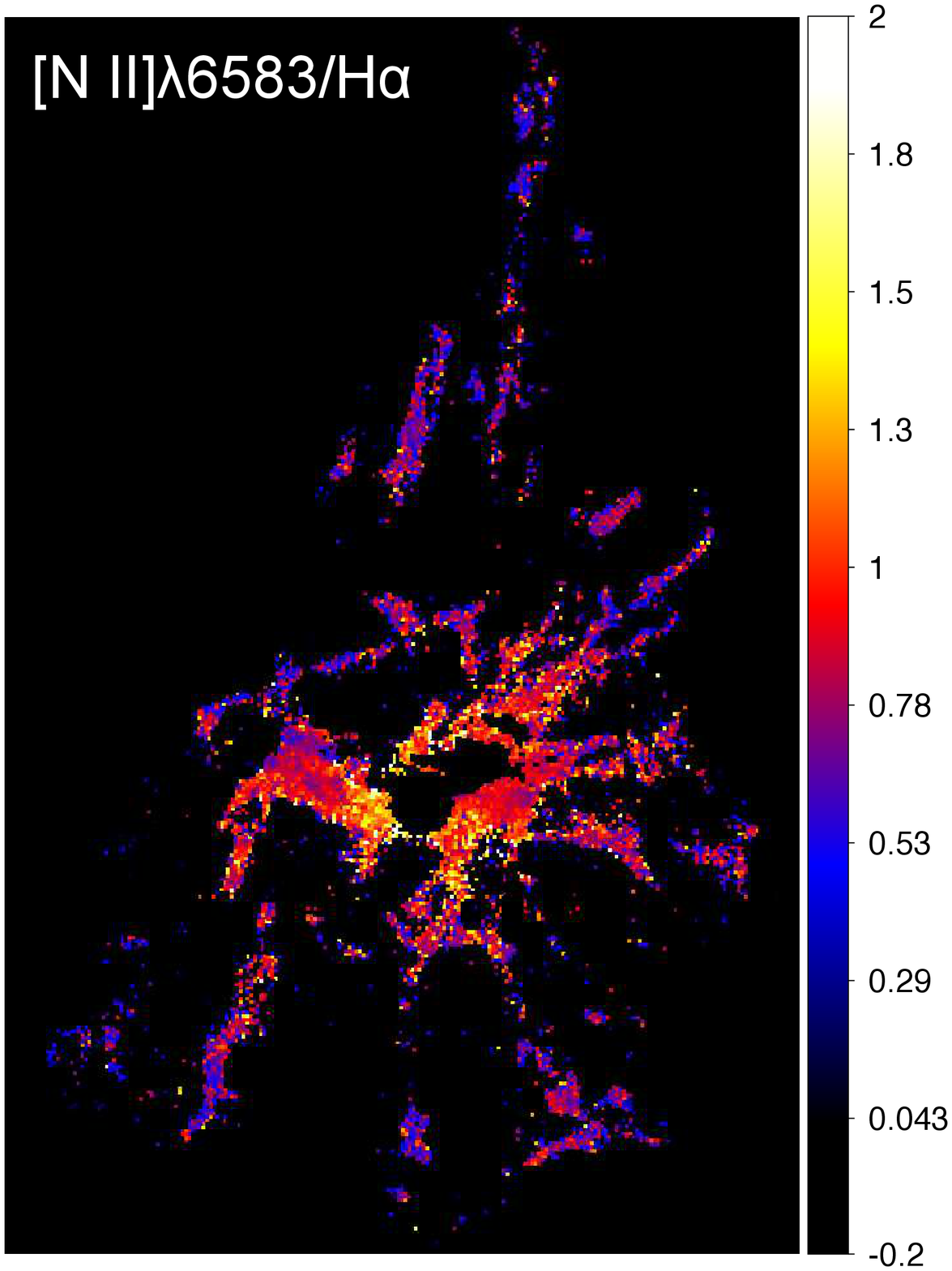}
    		\vspace{-10pt}
    \end{subfigure}
    \begin{subfigure}[t]{0.32\textwidth}
    		\centering    
    		\includegraphics[height=7.6cm]{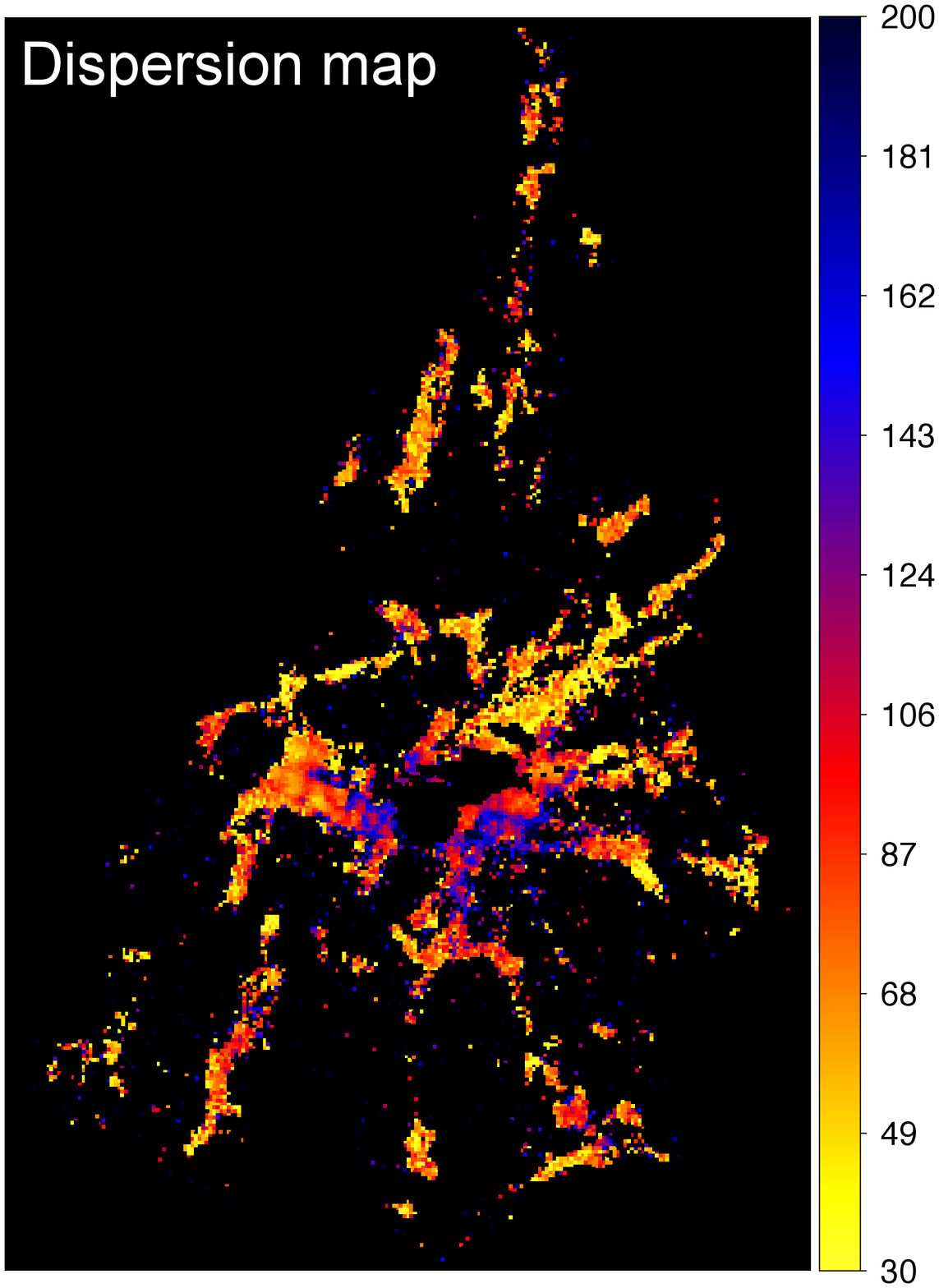}
    	    \vspace{-10pt}
    \end{subfigure} 
\caption{Flux map of $\text{H} \alpha$ emission in the LV system region (left, units are in erg/s/cm$^{2}$/pixel), [N II]$\lambda$6583/$\text{H} \alpha$ line ratio map (middle) and dispersion map (right, scale unit is in km/s). }
\label{fig:map_Ha_ratio_sigma}
\vspace{-17pt}
\end{figure*}

The complexity of this nebula arises from its several components: overlapping filaments with slightly different velocity shifts, the HV system and the AGN contribution. 
As we focus only on the LV component, pixels with [N II]$\lambda$6548, $\text{H} \alpha$ and [N II]$\lambda$6584 emission lines with a velocity shift close to the NGC 1275 systemic velocity were identified. 
The HV system was identified similarly, using a 8200 km/s systemic velocity, and subtracted. 
Fitting both the contribution from the AGN and the filaments, we found that the contribution from the AGN is predominant in terms of lines fluxes inside a radius of $6\arcsec$.
The central region centered at 3h19m48.1s + 41d30m42s with a radius of $6\arcsec$ was excluded. 
To increase the SNR without losing too much spatial resolution, we chose to bin the cube by a factor of 2.
The spectrum extracted from each binned remaining pixel was fitted using a Gaussian function convolved with the instrumental line shape - a sinc function \citep{martin_optimal_2016}.
The fitting software uses a least-squares Levenberg-Marquardt minimization algorithm \citep{levenberg_method_1944,marquardt_algorithm_1963} to fit the data.
We restricted the range of wavelengths to the band where [N II]$\lambda$6548, $\text{H} \alpha$ and [N II]$\lambda$6584 lines are found with the  systemic velocity shift of NGC 1275. 
Sky subtraction was done using the mean flux from a circular region with a radius of $20\arcsec$ centered at RA 03h19m58.57s and DEC $+41^{\circ}30\arcmin08.9\arcsec$, about $2\arcmin$ south-east of NGC 1275 nucleus.
The lines were fitted simultaneously, the velocity and broadening of the three lines grouped to reduce the number of parameters to fit. 
Only pixels with fitted $\text{H} \alpha$ flux higher than $30\times 10^{-18}$erg/s/cm$^{2}$/pixel were selected.
To directly compare our results with \cite{hitomi_collaboration_atmospheric_2017}, bulk velocities are calculated with respect to their redshift measurement: $v_{bulk} \equiv (z - 0.017284)*c_0 - 21.9  \text{km s}^{-1}$,  where $c_0$ is the speed of light and $- 21.9  \text{km s}^{-1}$ is the heliocentric correction based on the average value over the observation period from Astropy \textsc{SkyCoord.radial\char`_velocity\char`_correction()}.

\vspace{-20pt}
\section{Results and discussion}\label{Res}

\vspace{-5pt}
\subsection{Ionization mechanism}
\vspace{-5pt}

\begin{figure}
\vspace{-5pt}
\centering
\includegraphics[height=2.3in]{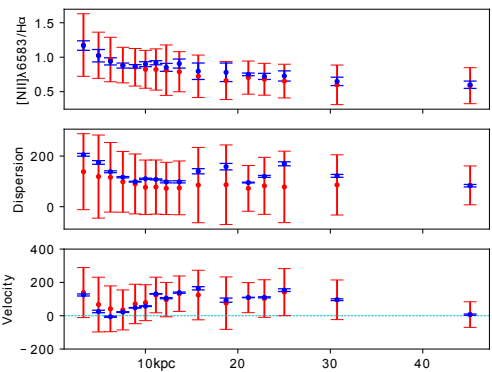}
\vspace{-10pt}
\caption{The mean (in red, with error bars indicating the standard deviation) and ensemble fit result (in blue) [N II]$\lambda$6583/$\text{H} \alpha$ line ratio, dispersion and velocity profiles taken in annuli containing 400 pixels centred on the AGN. Dispersions and velocities are given in km/s and the distance from the AGN is in kpc.}
\label{fig:Profiles}
\vspace{-15pt}
\end{figure}

Figure \ref{fig:map_Ha_ratio_sigma} (left) shows the $\text{H} \alpha$ flux map.
While the surface brightness of $\text{H} \alpha$ is mostly constant in the extended filaments, it is higher in the inner $\sim 30'' = 11$kpc.

Optical line ratios can be good indicators of the dominant excitation mechanism operating on the line-emitting gas (photoionization by stars, by a power-law continuum source or shock-wave heating, \citealt{baldwin_classification_1981}) .
The ratio [N II]$\lambda$6583/$\text{H} \alpha$ provides, for example, a measure of the ionization state of a gas.
When the source of ionization is stellar formation, this line ratio is a linear function of metallicity saturating at a value of $\sim 0.5$ for high metallicity (e.g. \citealt{kewley_host_2006}). 
The SITELLE [N II]$\lambda$6583/$\text{H} \alpha$ line ratio map of NGC 1275 is presented in figure \ref{fig:map_Ha_ratio_sigma} (middle) and the mean and ensemble ratio profiles taken in annuli containing 400 pixels is shown on figure \ref{fig:Profiles}. 
The line ratio varies through the map, being $\sim 0.5-1$ in the extended filaments, and above 1 in the central part of the nebula. 
However, streaks of star forming clusters associated with some filaments of NGC 1275 have been found \citep{canning_filamentary_2014}. 
Similar line ratio gradients have also been previously observed in the filaments of NGC 1275 \citep{hatch_origin_2006} and in several BCG with optical line emission (e.g. \citealt{hamer_optical_2016}).
The central region with higher line ratios could be related to energetic sources of ionization such as AGN and shocks, while filaments must be ionized by a source with lower power.
To effectively distinguished the source of ionization though, other line ratios are required, falling outside of the filter used during these observations.
The complete detailed BPT diagnostic of NGC 1275 nebula will be conducted using awarded SITELLE observations at 365-385 nm and 480-520 nm (PI: Gendron-Marsolais) and presented in future work (Gendron-Marsolais et al. in prep.).


\vspace{-15pt}
\subsection{Velocity dispersion measure across the nebula}
\vspace{-5pt}

According to the top-down multiphase condensation model, warm filaments and cold molecular clouds condensed out of the hot ICM through "chaotic cold accretion" (e.g. \citealt{gaspari_raining_2017}). This link between ICM and filaments imply that both must have the same ensemble velocity dispersion.
On the other hand, if these filaments are rather dragged out from the rise of AGN radio bubbles, they would be stabilized into the hot gas by magnetic fields, and therefore also sharing the same velocity field
\citep{fabian_magnetic_2008}.
The Hitomi Soft X-ray Spectrometer has shown that the line-of-sight velocity dispersions are on the order of $164\pm10$ km/s in the $30-60$ kpc region around the nucleus of the Perseus cluster \citep{hitomi_collaboration_quiescent_2016}, while SITELLE has provided an ensemble line-of-sight velocity dispersion of $137\pm20$ km/s \citep{gaspari_shaken_2017}.
In contrast, the dispersion map obtained from the fitting of each binned pixel shown in figure \ref{fig:map_Ha_ratio_sigma} shows instead smaller velocity dispersion in the filaments ($\lesssim 115$ km/s), but increasing linewidth closer to the center (up to $\sim 130$ km/s). SITELLE's level of resolution therefore probes smaller structures like individual filaments, rather than the ensemble multiphase gas.
Interestingly, the mean and ensemble dispersion profiles in figure \ref{fig:Profiles} show a general decrease up to $\sim 10$ kpc from the nucleus but a bump is visible between $\sim 15 - 20$ kpc. This corresponds to the region between the inner and ghost cavities and contains a known shock in the ICM to the north-east \citep{fabian_very_2006} which could be responsible for the higher mean dispersion.

\begin{figure}
\centering
\includegraphics[height=2.4in]{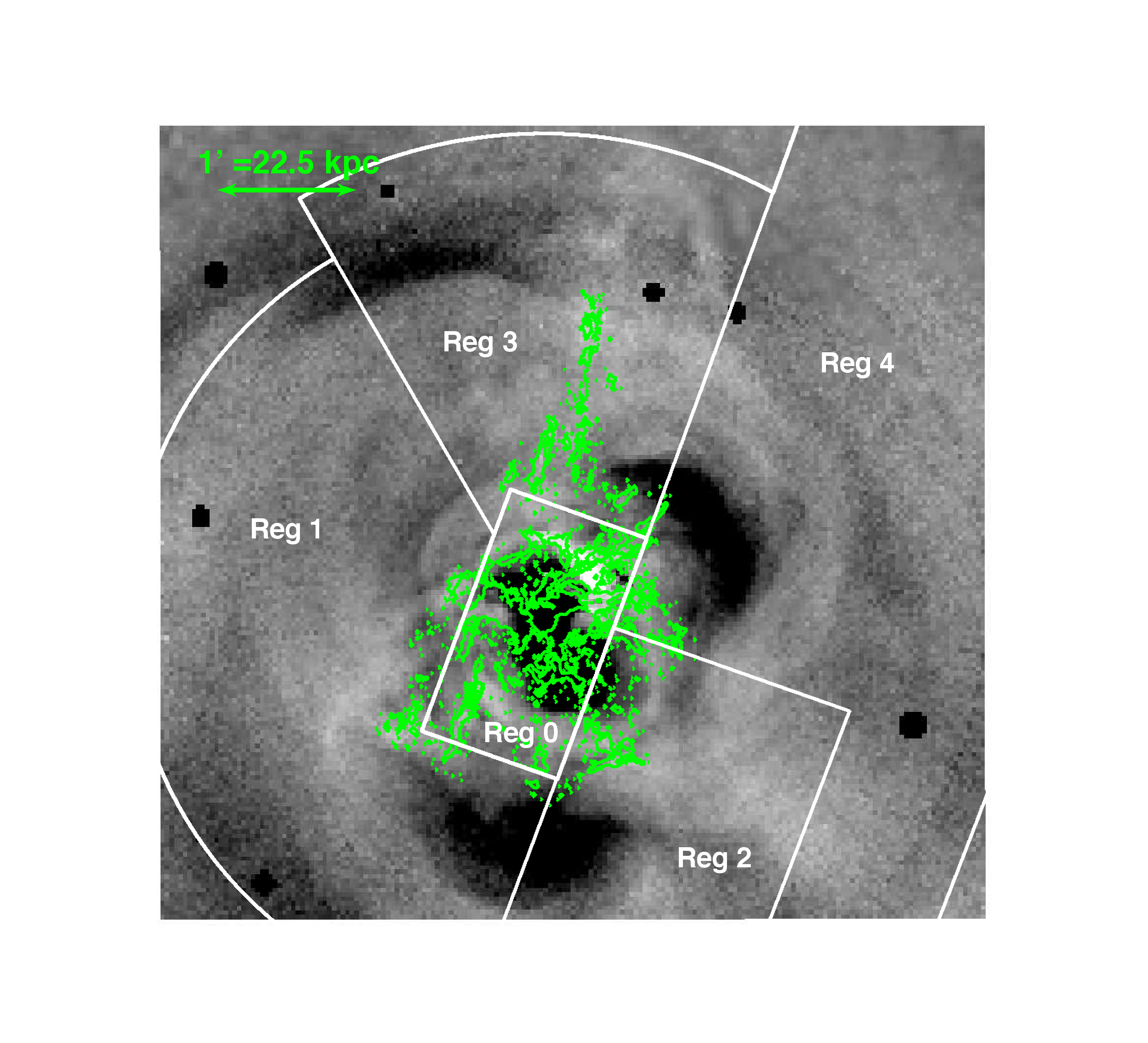}
\caption{\textit{Chandra} composite fractional residual image from \protect\cite{fabian_wide_2011} in the 0.5-7 keV band (1.4 Ms exposure). The PSF corrected Hitomi regions from  \protect\cite{hitomi_collaboration_atmospheric_2017} are shown in white. Contours from the LV system $\text{H} \alpha$ flux map (starting at $3\times 10^{-17}$ erg/s/cm$^{2}$/pixel) are shown in green.}
\label{fig:chandra_hitomi_reg}
\vspace{-15pt}
\end{figure}

Further comparisons can be explored between Hitomi and Sitelle line-of-sight velocity dispersions in the same regions as the ones used in \cite{hitomi_collaboration_atmospheric_2017} and shown in Figure \ref{fig:chandra_hitomi_reg}.
SITELLE pixels contained in each of those regions with $\text{H} \alpha$ and [N II] lines (including the AGN but excluding the HV system) were fitted as ensembles (see Table \ref{tab:hitomi}). The resulting velocities dispersions are as low ($\sim 120$ km/s) and uniform as the Hitomi measurements for the hot gas, supporting the infalling cooling flow model.  

\begin{table}
\centering
\caption{Comparison between Hitomi and Sitelle best-fitted bulk velocities and dispersions in regions shown on Figure \ref{fig:chandra_hitomi_reg}}
\label{tab:hitomi}
\vspace{-6pt}
\begin{tabular}{|c|c|c|c|c|}
\hline 
&\multicolumn{2}{c}{Hitomi} & \multicolumn{2}{c}{SITELLE}\\
Region & $v_{bulk}$ (km/s) & $\sigma_{v}$ (km/s) &  $v_{bulk}$ (km/s) &  $\sigma_{v}$ (km/s) \\ 
\hline 
Reg 0  &  $75_{-28}^{+26}$  &  $189_{-18}^{+19}$ & $48 \pm 3$   & $145 \pm 3$  \\[1ex]
Reg 1  &  $46_{-19}^{+19} $  &  $103_{-20}^{+19}$ & $-8 \pm 9$    & $155 \pm 9$  \\[1ex] 
Reg 2 &  $47_{-14}^{+14}$    &  $98_{-17}^{+17}$   &  $182 \pm 2$ & $116 \pm 3$  \\[1ex] 
Reg 3 &$-39_{-16}^{+15} $   &  $106_{-20}^{+20}$& $122 \pm 2$ & $94 \pm 2$ \\[1ex] 
Reg 4 &  $-77_{-28}^{+29}$ &  $218_{-21}^{+21}$ & $182 \pm 3$ & $88 \pm 3$  \\ 
\hline 
\end{tabular} 
\vspace{-6pt}
\end{table}

\begin{figure}
\centering
\includegraphics[height=3.7in]{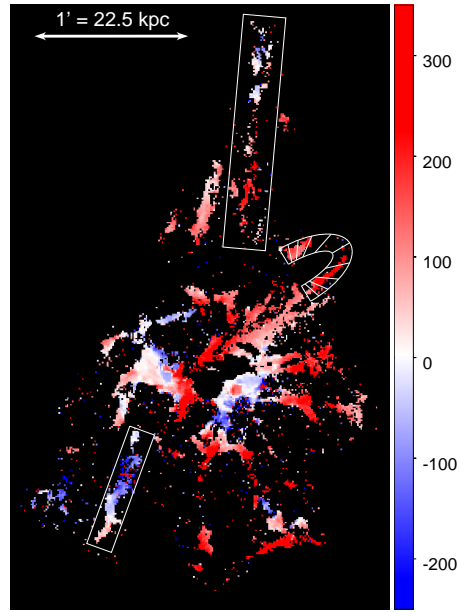}
\vspace{-7pt}
\caption{The velocity map of the LV system (in km/s). Profiles extracted from the three white regions are shown in figure \ref{fig:profile_fil}.}
\label{fig:vel_map}
\vspace{-15pt}
\end{figure}

\vspace{-15pt}
\subsection{Kinematics of the filaments}
\vspace{-5pt}

The velocity map of the LV system (see Figure \ref{fig:vel_map}) reveals a previously unknown rich velocity structure across the entire nebula. 
The presence of a larger scale velocity gradient is hard to extract from such a detailed map. 
We note that the median heliocentric velocity of the map is 5229 km/s, giving a high fraction of redshifted pixels ($\sim 80 \%$) relative to our chosen rest frame. We will discuss this difference in future work (Gendron-Marsolais et al. in prep.).
Overall, the mean and ensemble velocity profiles from figure \ref{fig:Profiles} do not show any clear radial gradient in velocity.
On average, the filaments do not seem to be falling smoothly and uniformly onto the galaxy nor do they seem to be pulled out of it.
No potential rotation, as suggested in \cite{conselice_nature_2001}, is visible.
The lack of ordered motion and the low measured velocities might indicate that these features are short lived, consistent with the molecular gas (e.g. \citealt{russell_alma_2016}).

\begin{figure}
\vspace{-5pt}
\centering
\includegraphics[height=3in]{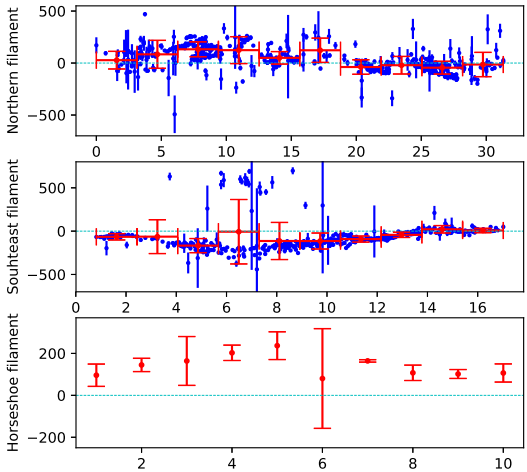}
\vspace{-10pt}
\caption{Velocity profiles extracted from the three white regions  on figure \ref{fig:vel_map} (in km/s). In the velocity profiles across the northern and the southeast filaments,  velocities from each pixel are in blue while the mean taken in ten bins of equal width are shown in red. These are plot against their distance from the base of the filaments (in kpc). A velocity profile across the horseshoe filament is also shown, the mean of the fitted velocities taken from ten bins.}
\label{fig:profile_fil}
\vspace{-15pt}
\end{figure}

With the spectral and spatial resolution provided by SITELLE, the kinematics of the filaments can be studied individually.
The line-of-sight velocity structure across the northern filament is very complex. Velocities from each pixel are plot against their distance from the AGN in figure \ref{fig:profile_fil} and show a scattered profile. The mean velocity profile taken in ten bins of equal width shows a more general trend, varying from positive to negative velocity as the radial distance to the nuclei increase. This is consistent with Gemini Multi-Object Spectrograph observations along this filament \citep{hatch_origin_2006}. The northern filament is therefore either stretching or collapsing depending on its de-projected orientation.
Furthermore, the southeast filament also shows complex dynamics with a mostly negative line-of-sight velocity, varying from $\sim - 100$ km/s to $-300$ km/s and increasing again up to  $\sim 0$ km/s  as the distance from the center increase (see figure \ref{fig:profile_fil}). 
Finally, the mean velocity profile across the horseshoe-shaped filament extracted from then bins is shown on figure \ref{fig:profile_fil}. 
It has an overall positive velocity increasing almost symmetrically on either side of the loop, reaching velocities of $\sim 200$ km/s. Again, this is similar to \cite{hatch_origin_2006} results and consistent with simulations of flow patterns below a rising bubble where the gas flows down on either sides of the bubble, the highest velocities located just behind the bubble.
Table \ref{tab:hitomi} shows the comparisons between Hitomi and Sitelle best-fitted bulk velocities.
Contrary to the velocity dispersions, we see no correlations between the bulk velocities of the warm and ionized gas, except for Reg 0.

\vspace{-25pt}
\section{Conclusion}\label{Conclusion}
\vspace{-5pt}
We have used SITELLE observations to probe the detailed dynamics of the filamentary nebula surrounding NGC 1275.
\vspace{-15pt}
\begin{enumerate}
\item We observe a smooth radial gradient of the [N II]$\lambda$6583/$\text{H} \alpha$ line ratio, suggesting a change in the ionization mechanism and source across the nebula: higher line ratios are found in the central region and must therefore be related to energetic sources of ionization (AGN and shocks), while filaments must be ionized by a source with lower power.
\item The velocity dispersions decrease with increasing distance from the center, but are as low and uniform as the Hitomi measurements of the ICM, while we see no correlations between the warm and the ionized gas bulk velocities.  
\item The velocity map of NGC 1275 revealed by SITELLE shows a previously unknown rich velocity structure across the entire nebula with no clear general trend or potential rotation, indicating that filaments are not falling uniformly onto the galaxy, nor being pulled out from it. 
\end{enumerate}
\vspace{-6pt}
These results demonstrate how SITELLE, with its large field of view, high angular and spectral resolution, is well suited for the study of emission-line nebulae among clusters' BCGs.

\vspace{-22pt}
\section*{Acknowledgments}
\vspace{-5pt}
MLGM is supported by the NSERC Postgraduate Scholarships-Doctoral Program.
JHL and LD are supported by NSERC through the discovery grant and Canada Research Chair programs, as well as FRQNT.
ACF acknowledges ERC Adanced Grant 340442.
ACE acknowledges support from STFC grant ST/P00541/1.
Based on observations obtained at the Canada-France-Hawaii Telescope (CFHT) which is operated from the summit of Maunakea by the National Research Council of Canada, the Institut National des Sciences de l'Univers of the Centre National de la Recherche Scientifique of France, and the University of Hawaii. The observations at the Canada-France-Hawaii Telescope were performed with care and respect from the summit of Maunakea which is a significant cultural and historic site.  


\vspace{-17pt}
\bibliographystyle{mnras}
\bibliography{biblio_perseus}

\begin{thebibliography}{}
\makeatletter
\relax
\def\mn@urlcharsother{\let\do\@makeother \do\$\do\&\do\#\do\^\do\_\do\%\do\~}
\def\mn@doi{\begingroup\mn@urlcharsother \@ifnextchar [ {\mn@doi@}
  {\mn@doi@[]}}
\def\mn@doi@[#1]#2{\def\@tempa{#1}\ifx\@tempa\@empty \href
  {http://dx.doi.org/#2} {doi:#2}\else \href {http://dx.doi.org/#2} {#1}\fi
  \endgroup}
\def\mn@eprint#1#2{\mn@eprint@#1:#2::\@nil}
\def\mn@eprint@arXiv#1{\href {http://arxiv.org/abs/#1} {{\tt arXiv:#1}}}
\def\mn@eprint@dblp#1{\href {http://dblp.uni-trier.de/rec/bibtex/#1.xml}
  {dblp:#1}}
\def\mn@eprint@#1:#2:#3:#4\@nil{\def\@tempa {#1}\def\@tempb {#2}\def\@tempc
  {#3}\ifx \@tempc \@empty \let \@tempc \@tempb \let \@tempb \@tempa \fi \ifx
  \@tempb \@empty \def\@tempb {arXiv}\fi \@ifundefined
  {mn@eprint@\@tempb}{\@tempb:\@tempc}{\expandafter \expandafter \csname
  mn@eprint@\@tempb\endcsname \expandafter{\@tempc}}}

\bibitem[\protect\citeauthoryear{Baldwin, Phillips  \& Terlevich}{Baldwin
  et~al.}{1981}]{baldwin_classification_1981}
Baldwin J.~A.,  Phillips M.~M.,   Terlevich R.,  1981, \mn@doi [PASP]
  {10.1086/130766}, 93, 5

\bibitem[\protect\citeauthoryear{Canning et~al.,}{Canning
  et~al.}{2014}]{canning_filamentary_2014}
Canning R. E.~A.,  et~al., 2014, \mn@doi [MNRAS] {10.1093/mnras/stu1191}, 444,
  336

\bibitem[\protect\citeauthoryear{Conselice, Gallagher  \& Wyse}{Conselice
  et~al.}{2001}]{conselice_nature_2001}
Conselice C.~J.,  Gallagher III J.~S.,   Wyse R. F.~G.,  2001, \mn@doi [ApJ]
  {10.1086/323534}, 122, 2281

\bibitem[\protect\citeauthoryear{Crawford, Allen, Ebeling, Edge  \&
  Fabian}{Crawford et~al.}{1999}]{crawford_rosat_1999}
Crawford C.~S.,  Allen S.~W.,  Ebeling H.,  Edge A.~C.,   Fabian A.~C.,  1999,
  \mn@doi [MNRAS] {10.1046/j.1365-8711.1999.02583.x}, 306, 857

\bibitem[\protect\citeauthoryear{Fabian, Sanders, Crawford, Conselice,
  Gallagher  \& Wyse}{Fabian et~al.}{2003}]{fabian_relationship_2003}
Fabian A.~C.,  Sanders J.~S.,  Crawford C.~S.,  Conselice C.~J.,  Gallagher
  J.~S.,   Wyse R. F.~G.,  2003, \mn@doi [MNRAS]
  {10.1046/j.1365-8711.2003.06856.x}, 344, L48

\bibitem[\protect\citeauthoryear{Fabian, Sanders, Taylor, Allen, Crawford,
  Johnstone  \& Iwasawa}{Fabian et~al.}{2006}]{fabian_very_2006}
Fabian A.~C.,  Sanders J.~S.,  Taylor G.~B.,  Allen S.~W.,  Crawford C.~S.,
  Johnstone R.~M.,   Iwasawa K.,  2006, \mn@doi [MNRAS]
  {10.1111/j.1365-2966.2005.09896.x}, 366, 417

\bibitem[\protect\citeauthoryear{Fabian, Johnstone, Sanders, Conselice,
  Crawford, Iii  \& Zweibel}{Fabian et~al.}{2008}]{fabian_magnetic_2008}
Fabian A.~C.,  Johnstone R.~M.,  Sanders J.~S.,  Conselice C.~J.,  Crawford
  C.~S.,  Iii J. S.~G.,   Zweibel E.,  2008, \mn@doi [Nature]
  {10.1038/nature07169}, 454, 968

\bibitem[\protect\citeauthoryear{Fabian et~al.,}{Fabian
  et~al.}{2011}]{fabian_wide_2011}
Fabian A.~C.,  et~al., 2011, \mn@doi [MNRAS]
  {10.1111/j.1365-2966.2011.19402.x}, 418, 2154

\bibitem[\protect\citeauthoryear{Forman, Kellogg, Gursky, Tananbaum  \&
  Giacconi}{Forman et~al.}{1972}]{forman_observations_1972}
Forman W.,  Kellogg E.,  Gursky H.,  Tananbaum H.,   Giacconi R.,  1972,
  \mn@doi [ApJ] {10.1086/151791}, 178, 309

\bibitem[\protect\citeauthoryear{Gaspari et~al.,}{Gaspari
  et~al.}{2017a}]{gaspari_shaken_2017}
Gaspari M.,  et~al., 2017a, accepted in ApJ, arXiv:1709.06564

\bibitem[\protect\citeauthoryear{Gaspari, Temi  \& Brighenti}{Gaspari
  et~al.}{2017b}]{gaspari_raining_2017}
Gaspari M.,  Temi P.,   Brighenti F.,  2017b, \mn@doi [MNRAS]
  {10.1093/mnras/stw3108}, 466, 677

\bibitem[\protect\citeauthoryear{Gendron-Marsolais et~al.,}{Gendron-Marsolais
  et~al.}{2017}]{gendron-marsolais_deep_2017}
Gendron-Marsolais M.,  et~al., 2017, \mn@doi [MNRAS] {10.1093/mnras/stx1042},
  469, 3872

\bibitem[\protect\citeauthoryear{Hamer et~al.,}{Hamer
  et~al.}{2016}]{hamer_optical_2016}
Hamer S.~L.,  et~al., 2016, \mn@doi [MNRAS] {10.1093/mnras/stw1054}, 460, 1758

\bibitem[\protect\citeauthoryear{Hatch, Crawford, Fabian  \& Johnstone}{Hatch
  et~al.}{2006}]{hatch_origin_2006}
Hatch N.~A.,  Crawford C.~S.,  Fabian A.~C.,   Johnstone R.~M.,  2006, \mn@doi
  [MNRAS] {10.1111/j.1365-2966.2006.09970.x}, 367, 433

\bibitem[\protect\citeauthoryear{{Hitomi Collaboration} et~al.,}{{Hitomi
  Collaboration} et~al.}{2017}]{hitomi_collaboration_atmospheric_2017}
{Hitomi Collaboration} et~al., 2017, arXiv:1711.00240

\bibitem[\protect\citeauthoryear{Hitomi~collaboration}{Hitomi~collaboration}{2016}]{hitomi_collaboration_quiescent_2016}
Hitomi~collaboration .,  2016, \mn@doi [Nature] {10.1038/nature18627}, 535, 117

\bibitem[\protect\citeauthoryear{Ho, Lim  \& {Dinh-V-Trung}}{Ho
  et~al.}{2009}]{ho_multiple_2009}
Ho I.-T.,  Lim J.,   {Dinh-V-Trung} 2009, \mn@doi [ApJ]
  {10.1088/0004-637X/698/2/1191}, 698, 1191

\bibitem[\protect\citeauthoryear{Kent \& Sargent}{Kent \&
  Sargent}{1979}]{kent_ionization_1979}
Kent S.~M.,  Sargent W. L.~W.,  1979, \mn@doi [ApJ] {10.1086/157125}, 230, 667

\bibitem[\protect\citeauthoryear{Kewley, Groves, Kauffmann  \& Heckman}{Kewley
  et~al.}{2006}]{kewley_host_2006}
Kewley L.~J.,  Groves B.,  Kauffmann G.,   Heckman T.,  2006, \mn@doi [MNRAS]
  {10.1111/j.1365-2966.2006.10859.x}, 372, 961

\bibitem[\protect\citeauthoryear{Levenberg}{Levenberg}{1944}]{levenberg_method_1944}
Levenberg K.,  1944, \mn@doi [Quart. Appl. Math.] {10.1090/qam/10666}, 2, 164

\bibitem[\protect\citeauthoryear{Lim, Ohyama, Chi-Hung, {Dinh-V-Trung}  \&
  Shiang-Yu}{Lim et~al.}{2012}]{lim_molecular_2012}
Lim J.,  Ohyama Y.,  Chi-Hung Y.,  {Dinh-V-Trung}  Shiang-Yu W.,  2012, \mn@doi
  [ApJ] {10.1088/0004-637X/744/2/112}, 744, 112

\bibitem[\protect\citeauthoryear{Lynds}{Lynds}{1970}]{lynds_improved_1970}
Lynds R.,  1970, \mn@doi [ApJL] {10.1086/180500}, 159

\bibitem[\protect\citeauthoryear{Marquardt}{Marquardt}{1963}]{marquardt_algorithm_1963}
Marquardt D.~W.,  1963, \mn@doi [SIAM J Appl Math] {10.1137/0111030}, 11, 431

\bibitem[\protect\citeauthoryear{Martin, Drissen  \& Joncas}{Martin
  et~al.}{2015}]{martin_orbs_2015}
Martin T.,  Drissen L.,   Joncas G.,  2015. ADASS XXIV, p.~327

\bibitem[\protect\citeauthoryear{Martin, Prunet  \& Drissen}{Martin
  et~al.}{2016}]{martin_optimal_2016}
Martin T.~B.,  Prunet S.,   Drissen L.,  2016, \mn@doi [MNRAS]
  {10.1093/mnras/stw2315}, 463, 4223

\bibitem[\protect\citeauthoryear{Martin, Drissen  \& Melchior}{Martin
  et~al.}{2018}]{martin_sitelle_2018}
Martin T.~B.,  Drissen L.,   Melchior A.-L.,  2018, MNRAS, 473, 4130

\bibitem[\protect\citeauthoryear{McDonald, Veilleux, Rupke  \&
  Mushotzky}{McDonald et~al.}{2010}]{mcdonald_origin_2010}
McDonald M.,  Veilleux S.,  Rupke D. S.~N.,   Mushotzky R.,  2010, \mn@doi
  [ApJ] {10.1088/0004-637X/721/2/1262}, 721, 1262

\bibitem[\protect\citeauthoryear{McDonald, Veilleux  \& Rupke}{McDonald
  et~al.}{2012}]{mcdonald_optical_2012}
McDonald M.,  Veilleux S.,   Rupke D. S.~N.,  2012, \mn@doi [ApJ]
  {10.1088/0004-637X/746/2/153}, 746, 153

\bibitem[\protect\citeauthoryear{Minkowski}{Minkowski}{1957}]{minkowski_optical_1957}
Minkowski R.,  1957. Proceedings 4th IAU Symposium, p.~107

\bibitem[\protect\citeauthoryear{Russell et~al.,}{Russell
  et~al.}{2016}]{russell_alma_2016}
Russell H.~R.,  et~al., 2016, \mn@doi [MNRAS] {10.1093/mnras/stw409}, 458, 3134

\bibitem[\protect\citeauthoryear{Salom{\'e} et~al.,}{Salom{\'e}
  et~al.}{2006}]{salome_cold_2006}
Salom{\'e} P.,  et~al., 2006, \mn@doi [A\&A] {10.1051/0004-6361:20054745}, 454,
  437

\bibitem[\protect\citeauthoryear{Salom{\'e}, Combes, Revaz, Downes, Edge  \&
  Fabian}{Salom{\'e} et~al.}{2011}]{salome_very_2011}
Salom{\'e} P.,  Combes F.,  Revaz Y.,  Downes D.,  Edge A.~C.,   Fabian A.~C.,
  2011, \mn@doi [A\&A] {10.1051/0004-6361/200811333}, 531, A85

\makeatother
\end{thebibliography}






\vspace{-9pt}
\bsp	
\label{lastpage}
\end{document}